\documentclass[aps,prb,twocolumn,notitlepage]{revtex4-2}

\usepackage{lipsum}
\usepackage{amsmath,graphicx,color}
\usepackage{bm}
\usepackage{natbib}
\usepackage{placeins}
\setcitestyle{square}

\begin{document}

\title{Emergence and Manipulation of non-equilibrium Yu-Shiba-Rusinov states}

\author{Jasmin Bedow, Eric Mascot and Dirk K. Morr}
\affiliation{University of Illinois at Chicago, Chicago, IL 60607, USA}

\date{\today}

\maketitle

{\bf The experimental advances in the study of time-dependent phenomena has opened a new path to investigating the complex electronic structure of strongly correlated and topological materials. Yu-Shiba-Rusinov (YSR) states induced by magnetic impurities in $s$-wave superconductors provide an ideal candidate system to study the response of a system to time-dependent manipulations of the magnetic environment. Here, we show that by imposing a time-dependent change in the magnetic exchange coupling, by changing the relative alignment of magnetic moments in an impurity dimer, or through a periodic drive of the impurity moment, one can tune the system through a time-dependent quantum phase transition, in which the system undergoes a transition from a singlet to a doublet ground state. We show that the electronic response of the system to external perturbations can be imaged through the time-dependent differential conductance, $dI(t)/dV$, which, in analogy to the equilibrium case, is proportional to a non-equilibrium local density of states. Our results open the path to visualizing the response of complex quantum systems to time-dependent external perturbations.
}

{\it Introduction} The experimental ability to probe and manipulate complex electronic correlations at the femtosecond time scale has opened unprecedented opportunities for the study of non-equilibrium quantum phenomena in strongly correlated or topological materials \cite{Perfetti2006,Schmitt2008,Fausti2011,Sobota2012,McIver2020}. The  development of next-generation spintronics and quantum computing applications requires the control of magnetic environments not only on similar time scales, but also on nanoscopic length scales. The latter has been achieved by using scanning tunneling spectroscopy (STS) techniques that have enabled the writing of magnetic skyrmions \cite{Romming2013}, the tuning of local magnetic exchange couplings \cite{Farinacci2018,Yang2019a,Karan2021}, and, in combination with electron spin resonance (ESR) techniques, the rotation of  individual magnetic moments in impurity clusters \cite{Yang2019}. These techniques were successfully applied \cite{Karan2021} to tune the energy of Yu-Shiba-Rusinov (YSR) states \cite{Yu1965,Shiba1968,Rusinov1969} -- induced by magnetic impurities placed on the surface of an $s$-wave superconductor -- and thus to drive the system through a quantum phase transition \cite{Sakurai1970,Salkola1997,Bazaliy2000,Balatsky2006}. While the required control of magnetic environments on electronic time scales has yet to be achieved, recent progress \cite{Yang2019} has raised the question not only of how quantum phenomena -- such as quantum phase transitions --  can be manipulated on the (electronic) femto- or picosecond time scale, but also of how such time-dependent phenomena can be described theoretically and visualized experimentally.

In this article, we address this question by investigating the manipulation of YSR states \cite{Yu1965,Shiba1968,Rusinov1969} on electronic time scales. By using two different theoretical methods based on the non-equilibrium Keldysh formalism \cite{Keldysh1965,Caroli1971,Rammer1986,Kamenev2009} we study the non-equilibrium emergence and manipulation of YSR states in response to external (STS and ESR based) perturbations of the magnetic environment. We show that the time evolution of YSR states can be visualized through the time-dependent differential conductance, $dI(t)/dV$, which, in analogy to the equilibrium case, is proportional to a non-equilibrium local density of states (LDOS), $N_{neq}$. These findings allow us to study the time-dependent phase transition of the system from a singlet to a doublet ground state by subjecting it to a time-dependent change in either the strength of a magnetic impurity’s exchange coupling, or in the relative orientation of moments in a magnetic dimer, or by externally driving a periodic precession of an impurity's magnetic moment. Moreover, we show that the extent to which the system is driven out-of-equilibrium is controlled by the time scale over which perturbations occur, and is directly reflected in the time and frequency dependence of $N_{neq}$. Finally, our formalism bridges the gap to periodically driven magnetic structures, allowing us to visualize the emergence of Floquet YSR states \cite{Kohler2005,Sacramento2015,Liu2019}. Our work thus provides a theoretical framework to study the emergence of non-equilibrium phenomena on electronic time and nanoscopic length scales in complex materials.\\

{\it Theoretical Method}
To study the non-equilibrium emergence and manipulation of YSR states, we consider magnetic impurities that are placed on the surface of an $s$-wave superconductor, with the exchange coupling or orientation of magnetic moments being time-dependent. Such a system is described by the Hamiltonian $\mathcal{H} = \mathcal{H}_0 + \mathcal{U}(t)$, where
\begin{align}
  \mathcal{H}_0 =& \, -t_e \sum_{\langle{\bf r} , {\bf r'}\rangle, \sigma} c^\dagger_{{\bf r},\sigma} c_{{\bf r'},\sigma} - \mu \sum_{\bf r} c^\dagger_{{\bf r}, \sigma} c_{{\bf r}, \sigma} \nonumber \\
  &+ \Delta_0 \sum_{{\bf r}} \left( c^\dagger_{{\bf r}, \uparrow} c^\dagger_{{\bf r}, \downarrow} + h.c. \right) \nonumber \; , \\
  \mathcal{U}(t) =& \, J(t) \sum_{{\bf r}, \alpha, \beta} c^\dagger_{{\bf r}, \alpha} \left[ {\bf S}_{\bf r}(t) \cdot {\bm \sigma} \right]_{\alpha \beta} c_{{\bf r}, \beta} \; ,
  \label{eq:H}
\end{align}
where $-t_e$ is the nearest-neighbor hopping parameter on a square lattice, $\mu$ is the chemical potential, $\Delta_0$ is the superconducting $s$-wave order parameter, and $c^\dagger_{\bf r, \sigma}$ creates an electron with spin $\sigma$ at site ${\bf r}$. $J$ is the magnetic exchange coupling and ${\bf S}_{\bf r}$ is the impurity spin at site ${\bf r}$ with the last sum running over all impurity positions.

To theoretically study the electronic response of the system to a time-dependent $J$ or ${\bf S}_{\bf r}$, we recall that the primary experimental probe in the study of YSR states is the differential conductance, $dI/dV$, measured in STS experiments \cite{Balatsky2006}. We therefore compute the time-dependent current flowing between an STS tip and the superconductor using a formulation of the non-equilibrium Keldysh formalism in the interaction representation \cite{Keldysh1965,Caroli1971,Rammer1986}, yielding
\begin{equation}
  I(V,t) = \frac{e}{\hbar} (-t_\text{tun}) \sum_\sigma \left[ G^{<}_{ts} (\sigma,t,t) - G^{<}_{st} (\sigma, t,t) \right] \; ,
  \label{eq:I}
\end{equation}
where $t_{\text{tun}}$ is the tunneling amplitude between the tip and the superconductor, $V$ is the applied bias, and $G_{ts}^{<}$ is the spin-dependent, equal-time lesser Green's function between the tip ($t$) and the site in the superconductor ($s$) that the electrons tunnel into \cite{Caroli1971}. To compute $G^{<}_{ts}$, we employ the Keldysh Dyson equations in real time, given by
\begin{align}
\bar{G}^{<}_{ts}(t,t) & = \bar{g}^{<}_{ts}(t,t) + \int dt_1 \bar{g}^{r}_{ts}(t,t_1) \bar{\mathcal{U}}(t_1) \bar{G}^{<}_{ss}(t_1,t) \nonumber \\
& + \int dt_1 \bar{g}^{<}_{ts}(t,t_1) \bar{\mathcal{U}}(t_1) \bar{G}^{a}_{ss}(t_1,t) \nonumber \\
\bar{G}^{a}_{ts}(t^\prime,t) & = \bar{g}^{a}_{ts}(t^\prime,t) + \int dt_1 \bar{g}^{a}_{ts}(t^\prime,t_1) \bar{\mathcal{U}}(t_1) \bar{G}^{a}_{ss}(t_1,t) \; ,
\label{eq:Dyson}
\end{align}
where $\bar{g}^{<,r,a}$ are the equilibrium lesser, retarded and advanced Greens function matrices of the unperturbed system
in Nambu space, and $\bar{\mathcal{U}}(t)$ is the time-dependent perturbation matrix arising from Eq.(\ref{eq:H}). By discretizing time, Eq.(\ref{eq:Dyson}) transforms into a set of coupled matrix equations, allowing us to obtain a closed form for $G^{<}_{ts}(t,t)$ (for details see Supplementary Note 1). Below all times are given in units of $\tau_e = \hbar /t_e$ which implies that for typical values of $t_e$ of a few hundred meV, $\tau_e$ is of the order of a few femtoseconds.

As the calculation of $dI/dV$ using Eq.(\ref{eq:I}) is computationally intensive, we also employ an alternative, computationally more efficient approach using the Heisenberg representation of the Keldysh formalism \cite{Rammer1986,Kamenev2009}.  This approach utilizes the fact that in equilibrium, the differential conductance is proportional to the local density of states $N({\bf r}, \omega) = -\mathrm{Im} \, g^r({\bf r,r},\omega)/\pi$. To define an analogous quantity out-of-equilibrium, we note that in the presence of a time-dependent perturbation, $G^r(t,t^\prime)$ depends explicitly on $t$ and $t^\prime$. Setting  $t^\prime = t - \Delta t$ and performing a Fourier transformation with regards to $\Delta t$, we obtain a time and frequency dependent Greens function $G^{r}(t, \omega)$, allowing us to define a ``non-equilibrium''
density of states, $N_{neq}$, via
\begin{align}\label{eq:NEQLDOS}
N_{neq}({\bf r}, \sigma, t, \omega) & = -\frac{1}{\pi} {\rm Im} G^{r}({\bf r},{\bf r}, \sigma, t, \omega) \; ,
\end{align}
where $G^{r}(t, \omega)$ is obtained from
\begin{equation}
  \left[ \mathrm{i} \frac{d}{dt} +  \omega  + i \Gamma - {\hat H}(t)\right] {\hat G}^r \left(t, \omega \right) = {\hat 1} \; ,
  \label{eq:diff_eq}
\end{equation}
for an arbitrary time dependence of the Hamiltonian. Here, ${\hat H}(t)$, ${\hat G}^r (t,\omega)$ and ${\hat 1}$ are matrices in real and Nambu space (for details, see Supplementary Note 2), and $\tau_\Gamma = \hbar/\Gamma$ is the quasi-particle lifetime.  We demonstrate below that $N_{neq}$ is proportional to the time-dependent $dI(V,t)/dV$, and thus represents a physical observable that describes the non-equilibrium time evolution of the superconductor's electronic structure. \\

{\it Results}
\begin{figure*}[t]
  \centering
  \includegraphics[width=17cm]{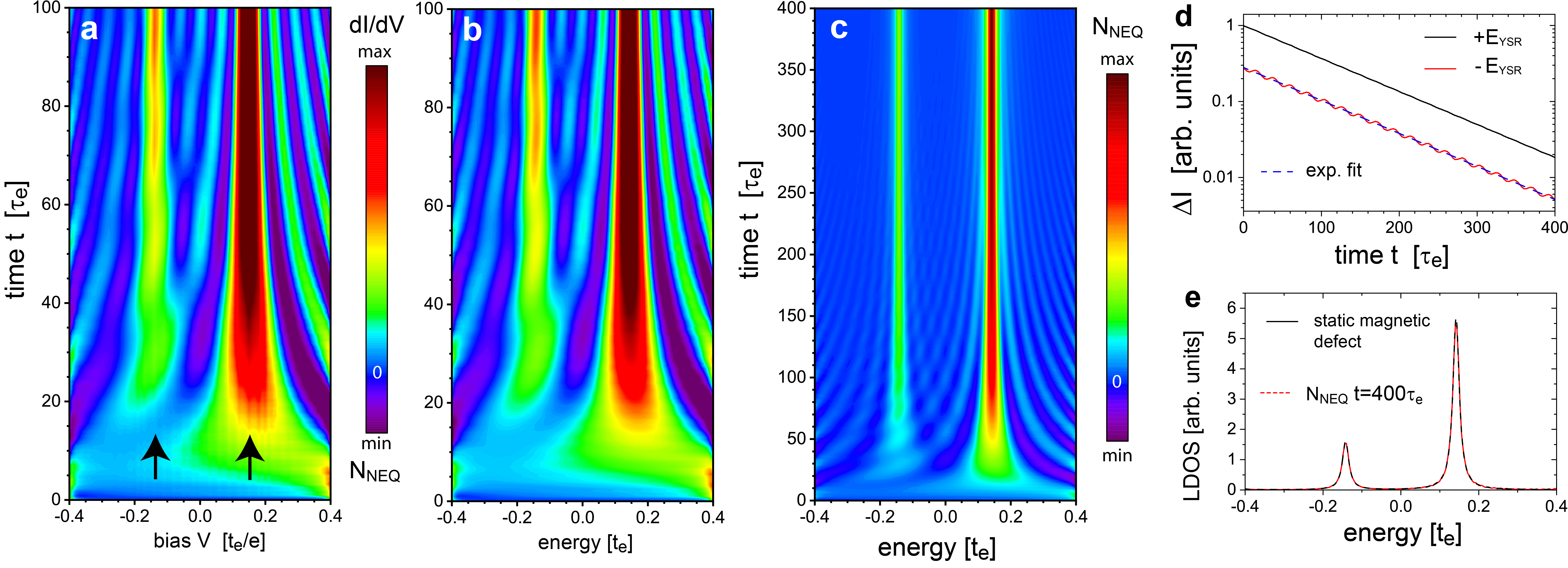}
  \caption{Time evolution of {\bf a} $dI/dV$, and {\bf b},{\bf c} $N_{neq}({\bf r}, t, \omega)$ at the site of the magnetic impurity. {\bf d} Time evolution of the deviation of the YSR peaks' height from equilibrium, $\Delta I$. {\bf e} Comparison of the equilibrium LDOS in the presence of a magnetic impurity, and $N_{neq}({\bf r}, t, \omega)$ at $t=400 \tau_e$. Here, and in the following figures $(\mu, JS, \Delta) = (-2.0, 1.8, 0.4)t_e$ and $\Gamma = 0.01 t_e$. The results for $N_{neq}$ were obtained for a real space system of size $50a_0 \times 50a_0$.}
  \label{fig:Fig1}
\end{figure*}
We begin by considering the question of how a YSR state emerges when a single magnetic impurity is placed on the surface of an $s$-wave superconductor at time $t=0$. To this end, we consider a perturbation of the form
\begin{equation}
\mathcal{U}(t \geq 0) =  JS \sum_{ \alpha, \beta} c^\dagger_{{\bf r}, \alpha} {\sigma}^z_{\alpha \beta} c_{{\bf r}, \beta} \; ,
\label{eq:pert1}
\end{equation}
and $\mathcal{U}(t<0)=0$. In Fig.~\ref{fig:Fig1}{\bf a}, we present the resulting time and bias dependence of $dI(V,t)/dV$ as obtained from Eq.(\ref{eq:I}) (due to the significant computational resources required,  the calculation of $dI/dV$ is limited to $t \leq 100 \tau_e$). At $t=0$, the LDOS is that of an unperturbed $s$-wave superconductor without a magnetic impurity, exhibiting a hard gap and coherence peaks at $\pm \Delta_0$. As the magnetic impurity is placed on the surface, spectroscopic weight from the gap edges begins to be transferred into the gap, creating at first a broad peak centered around the energy of the emerging YSR state (see black arrows in Fig.~\ref{fig:Fig1}{\bf a}). This shift of spectroscopic weight occurs in a wave-like form, as reflected in the stripe-like patterns of $dI(V,t)/dV$ in Fig.~\ref{fig:Fig1}{\bf a}.  With increasing time, the width of the peaks decreases, while their height increases, which is a direct consequence of the uncertainty principle: $t \Delta E \geq \hbar$ implies that as the time over which the impurity is located on the surface increases, the YSR state's energy uncertainty, i.e., the peaks' width decreases.

To elucidate the relation between $dI/dV$ and $N_{neq}$, we note that for the perturbation of Eq.(\ref{eq:pert1}), the solution of Eq.(\ref{eq:diff_eq}) is given by
\begin{align}
    \hat{G}^r(t,\omega) =& \left[1 - e^{\mathrm{i} \left( \omega \hat{1} - \hat{H}^{+} \right) t} e^{ - \Gamma t} \right] \hat{g}^r_+(\omega) \nonumber \\
    & + e^{\mathrm{i} \left( \omega \hat{1} - \hat{H}^{+} \right) t} e^{ - \Gamma t} \hat{g}^r_0(\omega)
    \label{eq:G_analytic}
\end{align}
where ${\hat g}^{r}_{0,+}$ are the equilibrium Green's functions for a system without and with an impurity, respectively, and $\hat{H}^+$ is the full Hamiltonian for $t \geq 0$. We thus have ${\hat G}^r(t=0, \omega)={\hat g}^{r}_{0} (\omega)$ and ${\hat G}^r(t\rightarrow \infty, \omega)={\hat g}^{r}_{+} (\omega)$.  A comparison of $N_{neq}$ obtained from Eqs.(\ref{eq:NEQLDOS}) and (\ref{eq:G_analytic}) (see Fig.~\ref{fig:Fig1}{\bf b}) with $dI/dV$ (see Fig.~\ref{fig:Fig1}{\bf a}) shows remarkable quantitative agreement (up to an overall scaling factor) in the $(\omega,t)$-plane. This suggests that $dI(V,t)/dV \sim N_{neq}(t, \omega = eV)$, in analogy to their equilibrium counterparts. We thus conclude that $N_{neq}$ represents a physical observable that describes the non-equilibrium time evolution of the superconductor's electronic structure. As the calculation of $N_{neq}$ is computationally less demanding than that of $dI/dV$, we can now study the time-evolution  of the system up to much larger times using $N_{neq}$, as shown in
Fig.~\ref{fig:Fig1}{\bf c}, where we present $N_{neq}(t, \omega)$ up to $t=400 \tau_e$. Eq.(\ref{eq:G_analytic}) suggests that when $\omega=\pm E_{YSR}$, with $E_{YSR}$ being the YSR state energy, the oscillatory term, $\sim e^{i\left(\omega {\hat 1} - {\hat H^+}\right)t}$ is identical to one, and $N_{neq}$ relaxes exponentially to the static case for $t \rightarrow \infty$, with the relaxation time given by $\tau_\Gamma$. This is confirmed by a log-plot of the deviation of the YSR peaks' height from the equilibrium value, $\Delta I$, as a function of time presented in Fig.~\ref{fig:Fig1}{\bf d}, where the dashed line corresponds to $\sim e^{-\Gamma t}$.  Finally, a comparison of the equilibrium LDOS in the presence of a static magnetic impurity with $N_{neq}(t, \omega)$ at $t=400 \tau_e$ (see Fig.~\ref{fig:Fig1}{\bf e}) shows very good agreement, demonstrating that this formalism allows us to study the system during the entire equilibration process.

\begin{figure}
  \centering
  \includegraphics[width=\columnwidth]{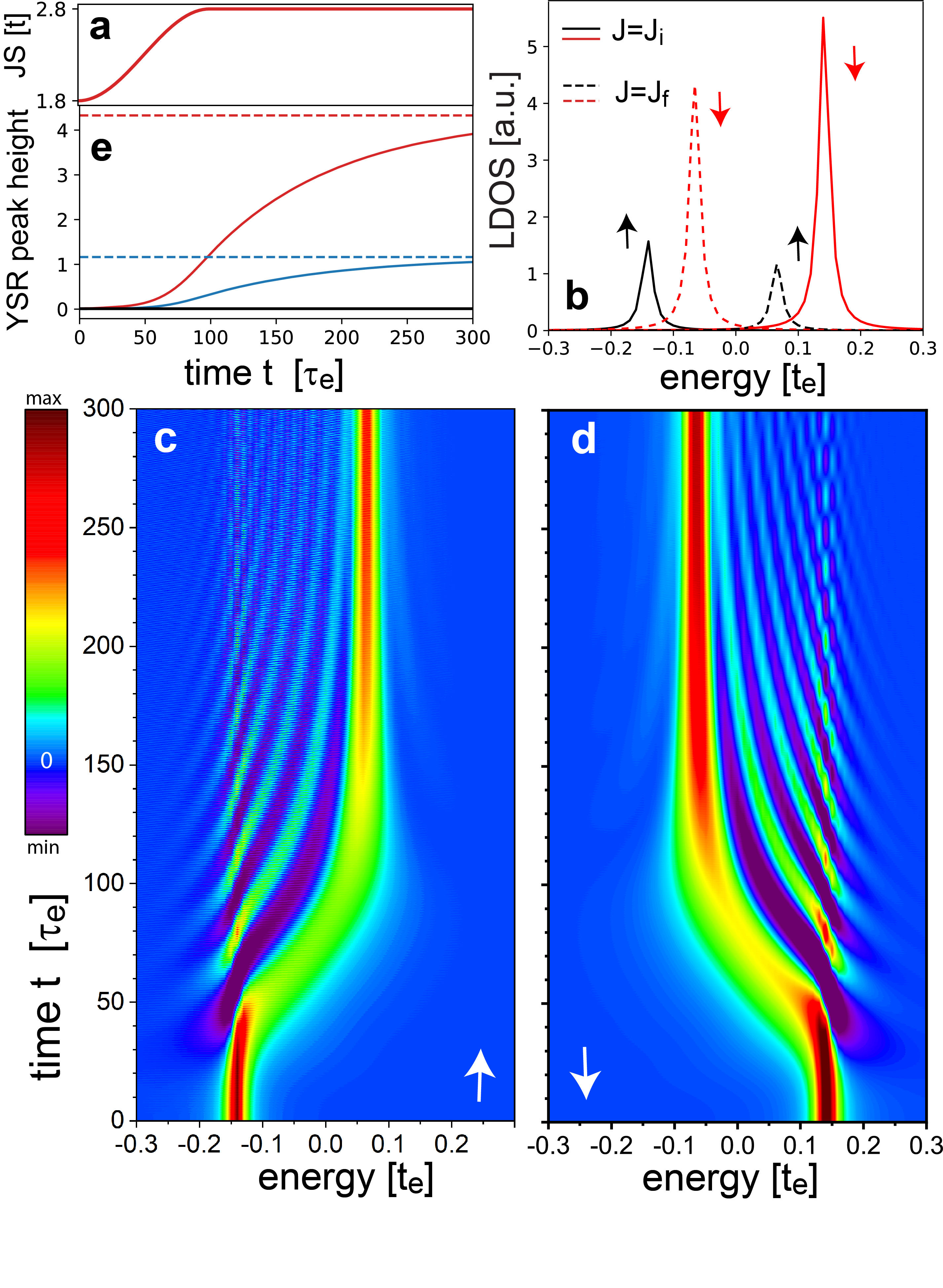}
  \caption{{\bf a} Time dependence of $J(t)$ with $J_i = 1.8t_e<J_c$, $J_f = 2.8t_e>J_c=2.45t_e$) and $\Delta t = 100 \tau_e$. {\bf b} Spin-resolved equilibrium LDOS in the presence of a single magnetic impurity with $J_i<J_c$ and $J_f>J_c$. Time dependence of the spin-resolved $N_{neq}(\sigma, t, \omega)$ for {\bf c} $\sigma=\uparrow$, and {\bf d} $\sigma=\downarrow$ arising from $J(t)$ in panel {\bf a}. {\bf e} Height of the YSR peaks as a function of time, representing the relaxation of the system.}
  \label{fig:Fig3}
\end{figure}
An intriguing phenomenon associated with the presence of a magnetic impurity is the possibility to drive the superconductor through a quantum phase transition between a singlet and doublet ground state by either tuning $J$ through a critical value, $J_c$, \cite{Sakurai1970,Salkola1997,Bazaliy2000,Balatsky2006} or by changing the distance or relative spin alignment of two or more impurities \cite{Morr2003,Morr2006}. This phase transition is accompanied by a zero-energy crossing of the YSR peaks in the LDOS. To investigate whether such a phase transition can also be induced using non-equilibrium techniques, we begin by considering the case of a time-dependent exchange coupling $J$, which is motivated by the experimental ability to significantly vary $J$ using STS and ESR techniques \cite{Farinacci2018,Yang2019a,Karan2021}. Specifically, we increase the exchange coupling from an initial value $J_i <J_c$ at $t=0$ to a final value $J_f >J_c$ at a time $t=\Delta t$, using the time evolution $J(t) = J_i + (J_f-J_i) \sin^2\left(\frac{\pi t}{2\Delta t} \right)$ for  $0 \leq t \leq \Delta t$ as shown in Fig.~\ref{fig:Fig3}{\bf a}. A plot of the equilibrium LDOS for $J=J_{i,f}$ in Fig.~\ref{fig:Fig3}{\bf b} demonstrates that the zero energy crossing of the YSR peaks at the phase transition exchanges the spin projection of their particle- and hole-like branches  \cite{Sakurai1970,Salkola1997,Bazaliy2000,Balatsky2006}, which can be mapped using spin-polarized STS \cite{Cornils2017}. The time evolution of the phase transition can thus be best visualized by considering the spin-resolved $N_{neq}(\sigma, t, \omega)$, shown in Figs.~\ref{fig:Fig3}{\bf c} and {\bf d} for $\sigma=\,\uparrow$ and $\sigma=\,\downarrow$, respectively. While the change in $J$ over a finite time $\Delta t$ leads to significant oscillations in $N_{neq}$, a zero-energy crossing of the YSR peaks, indicating a time-dependent phase transition \cite{Sakurai1970,Salkola1997,Bazaliy2000,Balatsky2006}, can still be identified. Moreover, $N_{neq}$ reveals several noteworthy features. First, the spectral weight of the YSR peaks is  decreased during the zero-energy crossing, concomitant with the phase transition. Second, a substantial redistribution of the YSR state's spectral weight begins to occur only after  $t \approx 50 \tau_e = \Delta t/2$, indicating a delayed response of the system to the imposed change in $J$. This is particularly evident from comparison of the time-dependent YSR peak height at the final energy position in Fig.~\ref{fig:Fig3}{\bf e} with $J(t)$ shown in Fig.~\ref{fig:Fig3}{\bf a}. Third, substantial oscillations in $N_{neq}$ at the energy of the original YSR peaks persist up to time scales significantly longer than $\Delta t$, and the spectral weight of these oscillations is shifted to the final energy position of the YSR states in a wave-like pattern, as evidenced by the stripe-like pattern in $N_{neq}$ (see Figs.~\ref{fig:Fig3}{\bf c} and {\bf d}).

\begin{figure}
  \centering
  \includegraphics[width=\columnwidth]{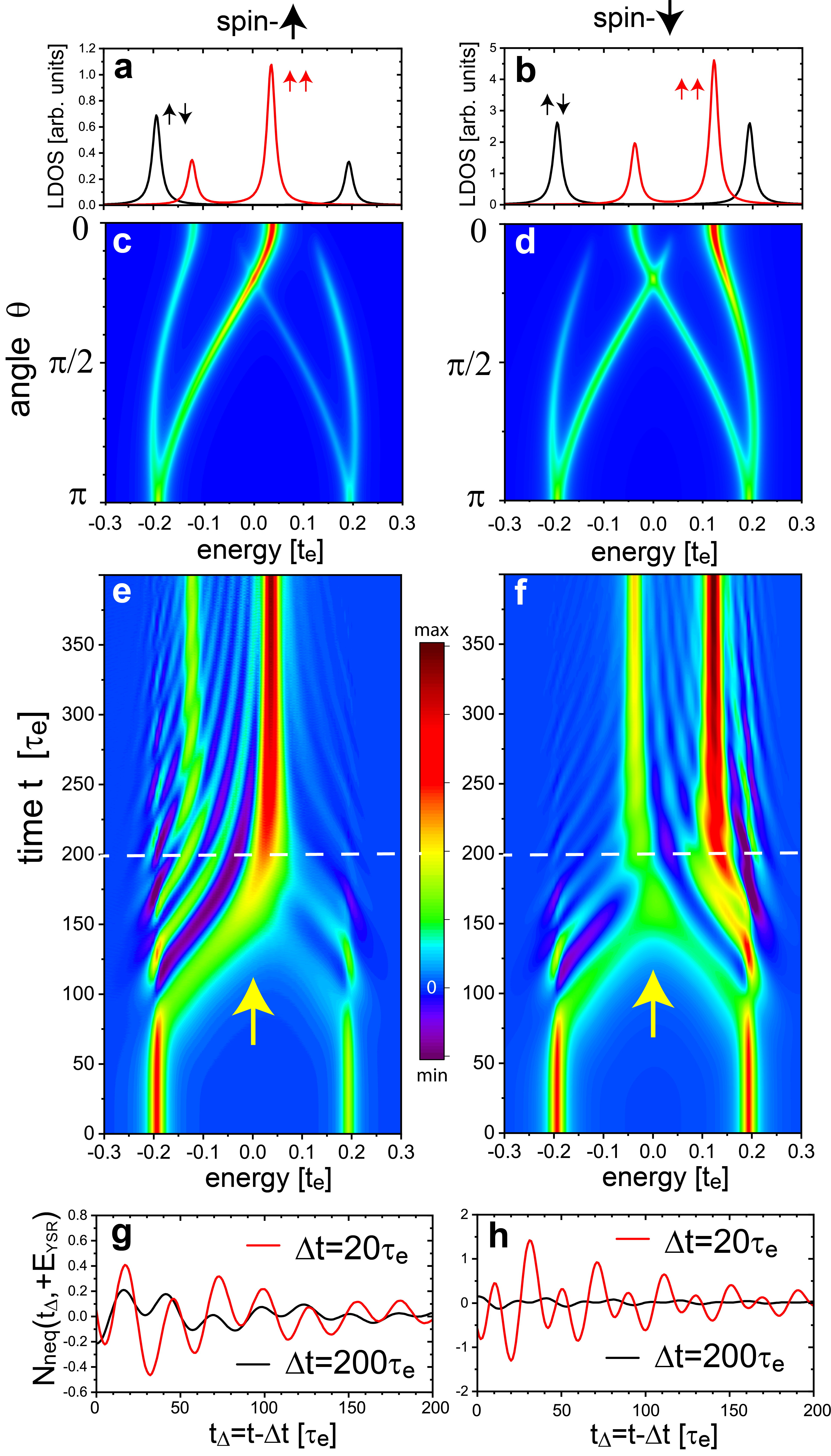}
  \caption{Equilibrium {\bf a} spin-$\uparrow$ and {\bf b} spin-$\downarrow$ LDOS for parallel ($\uparrow\uparrow$) and antiparallel ($\uparrow \downarrow$) alignment of the spins of two magnetic impurities, located on nearest-neighbor sites. Quasi-static evolution of the {\bf c} spin-$\uparrow$, {\bf d} spin-$\downarrow$ LDOS as a function of the polar angle $\theta$. Spin-resolved $N_{neq}(\sigma, t, \omega)$ for {\bf e} $\sigma=\uparrow$, and {\bf f} $\sigma=\downarrow$ arising from the relative rotation of the impurity moments. $N_{neq}(\sigma, t_\Delta, \omega)$ for {\bf g} $\sigma=\uparrow$, and {\bf h} $\sigma=\downarrow$ (see text).}
  \label{fig:Fig2}
\end{figure}
The superconductor can also be tuned through a quantum phase transition by changing the relative alignment of the spins of two impurities in close proximity from antiparallel to parallel \cite{Morr2003}, even if each of the two impurities possesses an exchange coupling smaller than the critical value, $J_c$. Such a change in the spins' relative alignment in an impurity dimer has been achieved using ESR techniques \cite{Yang2019}. For antiparallel alignment, the YSR states associated with each impurity cannot hybridize as the spin quantum number of their particle- and hole-like branches are opposite. The LDOS thus exhibits two pairs of degenerate YSR states with opposite spin quantum numbers \cite{Morr2003}, as shown in Figs.~\ref{fig:Fig2}{\bf a} and {\bf b}. When the alignment is rotated to parallel, the YSR states hybridize, leading to an energy splitting between them and the emergence of four peaks in the LDOS (see Figs.~\ref{fig:Fig2}{\bf a},{\bf b}). Plotting the equilibrium LDOS as a function of the angle $\theta$ between the spins (Figs.~\ref{fig:Fig2}{\bf c},{\bf d}), corresponding to the adiabatic limit of the rotation, we find that the energy splitting increases with decreasing $\theta$, leading to a zero energy crossing of two of the YSR peaks, and a concomitant phase transition from a singlet to a doublet ground state. The question naturally arises as to the time dependence of such a phase transition when the relative alignment is changed over a finite time $\Delta t$, as was done in Ref.~\cite{Yang2019} through the application of an ESR $\pi$ pulse.
To address this question, we consider for concreteness the rotation of one of the two impurity spins through the $xz$ plane described by ${\bf S}_{\bf r} = S[\sin{\theta(t)},0,\cos{\theta(t)}]$, with the time-dependent polar angle given by
\begin{align}
  \theta(t) &=
    \pi \cos^2\left(\frac{\pi}{2\Delta t} t \right), \quad {\rm for} \  0 \leq t \leq \Delta t \ .
\end{align}
In Figs.~\ref{fig:Fig2}{\bf e},{\bf f} we present the resulting spin-resolved $N_{neq}(\sigma, t, \omega)$ for $\Delta t = 200 \tau_e$, with the spins being antiparallel for $t=0$ and parallel for $t = \Delta t$. As before, we can identify a time-dependent zero-energy crossing of the YSR peaks  (see yellow arrows in Figs.~\ref{fig:Fig2}{\bf e},{\bf f}) indicating a phase transition from a singlet to a doublet ground state. We note that the angular dependence of the equilibrium LDOS shown in Figs.~\ref{fig:Fig2}{\bf c},{\bf d}, corresponds to the time dependence of $N_{neq}$ between $0 \leq t \leq \Delta t$ (see white dashed line in Figs.~\ref{fig:Fig2}{\bf e},{\bf f}) in the adiabatic limit of $\Delta t \rightarrow \infty$. A comparison of the equilibrium LDOS with $N_{neq}$ in Figs.~\ref{fig:Fig2}{\bf e},{\bf f} thus reveals that a finite $\Delta t$ leads to significant time-dependent oscillations in $N_{neq}$, in particular near the energy positions of the YSR peaks at $t=0$, $\pm E_{YSR}^{(0)}$. To investigate how the strength of these oscillations depends on $\Delta t$, we plot in Figs.~\ref{fig:Fig2}{\bf g} and {\bf h} $N_{neq}(t_\Delta, \pm E_{YSR}^{(0)})$ as a function of time after the rotation is completed, i.e., $t_\Delta = t-\Delta t$, for two different values of $\Delta t$. We find that the amplitude of the oscillations in $N_{neq}(t_\Delta, \omega)$ increases with decreasing $\Delta t$ as the system is more strongly driven out of equilibrium. This result is consistent with the expectation that in the quasi-static, adiabatic limit, $\Delta t \rightarrow \infty$, these oscillations vanish, and $N_{neq}$ becomes identical to the equilibrium LDOS in Figs.~\ref{fig:Fig2}{\bf c} and {\bf d}, when plotted as a function of angle.

\begin{figure}
  \centering
  \includegraphics[width=\columnwidth]{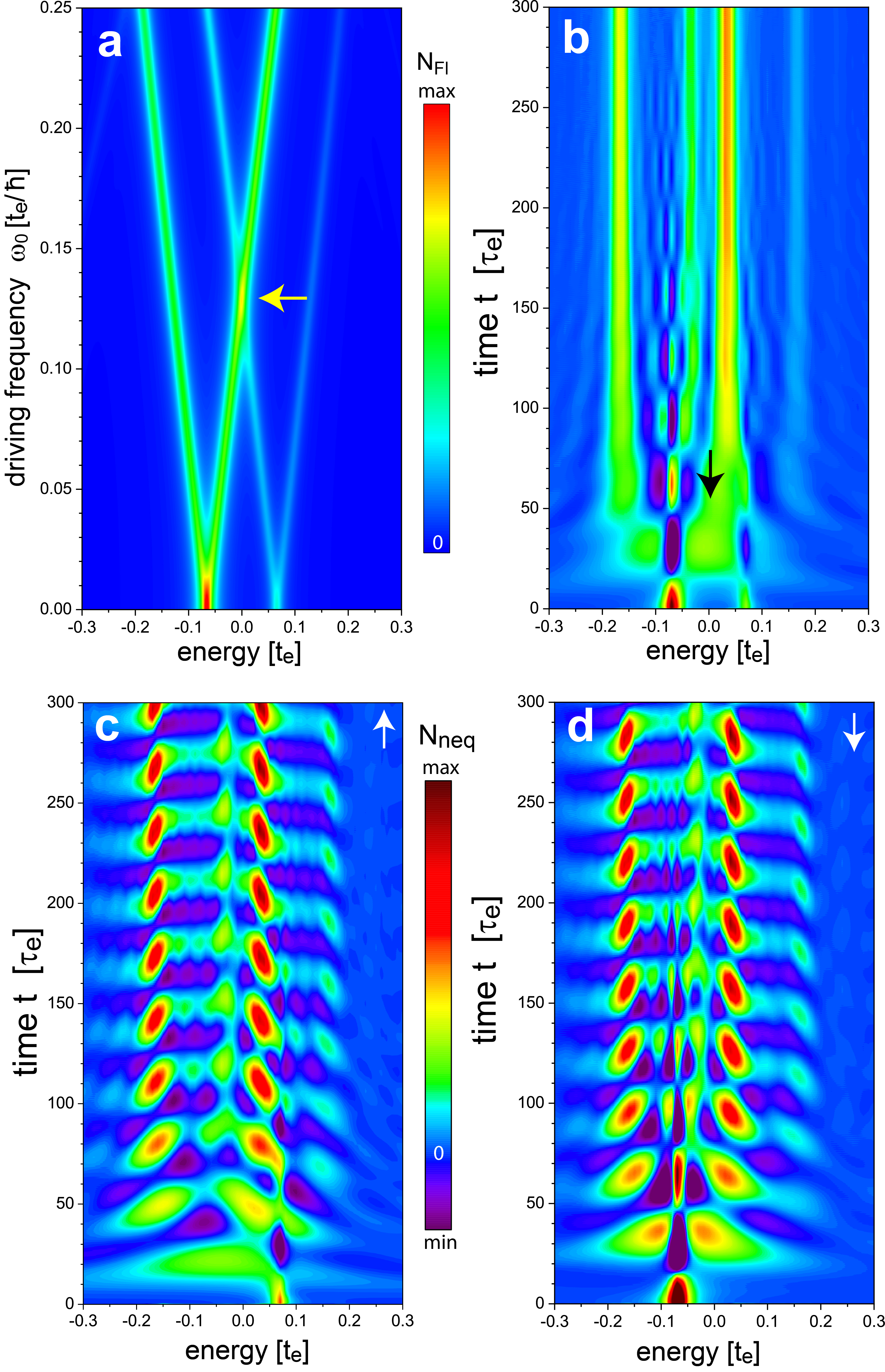}
  \caption{ {\bf a} Floquet LDOS $N_{Fl}$ as a function of energy for a continuously rotating spin with driving frequency $\omega_0$, with $l=2$ as the cutoff in the Floquet-Sambe matrix. Time evolution of {\bf b} $N_{neq}(t, \omega)$,  {\bf c} $N_{neq}(\uparrow, t,  \omega)$ and {\bf d} $N_{neq}(\downarrow, t,  \omega)$ for $\omega_0 = 0.2 t_e/\hbar$, corresponding to $T=10 \pi \tau_e$. Here, $J=2.8 t_e > J_c$. }
  \label{fig:Fig4}
\end{figure}

Finally, to bridge the gap to periodically driven systems that can be described using the Floquet formalism \cite{Kohler2005,Sacramento2015,Liu2019}, we consider the time evolution from a static impurity spin at $t=0$ to an impurity spin that rotates periodically in the $xz$-plane with a driving frequency of $\omega_0 = 2\pi/T$. We note that the periodic rotation of the impurity's spin leads to a splitting of the (static) YSR peaks in the density of states \cite{Kaladzhyan2017}. This is shown in Fig.~\ref{fig:Fig4}{\bf a} where we present the evolution of the Floquet LDOS $N_{Fl}$ with increasing driving frequency $\omega_0$ (for details, see Supplementary Note 3). Since, $J=2.8 t_e > J_c$, and since $J$ is effectively reduced with increasing $\omega_0$, a zero-energy crossing of the YSR peaks in $N_{Fl}$ occurs at some critical value of the driving frequency, $\omega_c$, indicating a phase transition (see yellow arrow in Fig.~\ref{fig:Fig4}{\bf a}), this time from a doublet to a singlet ground state. We note that for $\omega \rightarrow \infty$ the YSR states merge with the continuum, as the scattered electrons see a vanishing effective $J$. To investigate the transition from a static to a periodically rotating impurity spin, we employ a time-dependent polar angle given by
$\theta(t) = \omega_0 t \tanh\left(5\omega_0t/\pi \right)$ for $0 \leq t \le T/2 $
and $\theta(t) =\omega_0 t$ for $t \geq T/2$. The resulting $N_{neq}$ shown in Fig.~\ref{fig:Fig4}{\bf b} for $\omega_0 = 0.2 t_e /\hbar$ clearly reveals the emergence of the YSR peak splitting and of four YSR peaks with increasing time. Since $\omega_0 > \omega_c$ and $J>J_c$, we find that with increasing time, there is a substantial transfer of spectral weight across zero energy, indicating a time-dependent phase transition (see black arrow in Fig.~\ref{fig:Fig4}{\bf b}). We note that for $t \gg T$, the system is described by a Hamiltonian that is periodic in time, such that $N_{neq}$ is expected to coincide with the Floquet LDOS $N_{Fl}$. We find that already for $t=400 \tau_e$, $N_{neq}$ is in very good quantitative agreement with $N_{Fl}$ (see Supplementary Note 4) implying that this formalism is suited to describe not only the transition from a static to a periodically driven system, but also a periodically driven system itself. Moreover, we note that while the total, spin-summed $N_{neq}(t, \omega)$ does not exhibit any signature of the rotation period (consistent with a time-independent total $N_{Fl}$), the spin-resolved $N_{neq}(\sigma, t, \omega)$ shown in Figs.~\ref{fig:Fig4}{\bf }c and {\bf d} does. As expected from a rotation of the spin in the $xz$-plane, we find that $N_{neq}$ for $\sigma=\uparrow$ and $\sigma=\downarrow$ are out of phase, i.e., shifted by half a period $T$.
\\

{\it Discussion} The ability to understand the non-equilibrium response of complex quantum systems to external perturbations on electronic (femto-second) timescales and nanoscopic length scales is crucial not only for advancing our theoretical understanding of these systems, but also for the development of next generation spintronics and quantum computing applications. While ARPES experiments have probed complex systems at the femtosecond scale \cite{Perfetti2006,Schmitt2008,Fausti2011,Sobota2012,McIver2020}, it is presently unclear whether STS experiments will ever achieve electronic time scale resolution, which would be necessary to image the non-equilibrium quantum phenomena discussed above. However, recent results suggest that progress in increasing the time resolution might be made by combining STS and ESR techniques \cite{Yang2019}. Here, we have shown that time-dependent manipulations of magnetic impurities -- leading to changes in their exchange coupling or relative spin alignment --  not only allow us to study the emergence of quantum phenomena, such as the formation of YSR states, but can also give rise to new intriguing phenomena, such as a time-dependent phase transition. Indeed, studying and understanding the response of complex quantum system to external perturbations will be crucial for realizing topological quantum gates, and the required braiding of Majorana fermions. The formalism we have presented here provides a suitable approach for achieving this objective. \\

\noindent{\bf Acknowledgments}\\
The authors would like to thank S. Rachel, J. Wiebe, and R. Wiesendanger for stimulating discussions.\\

\noindent{\bf Funding}\\

 This work was supported by the U. S. Department of Energy, Office of Science, Basic Energy Sciences, under Award No. DE-FG02-05ER46225.\\

\end{document}


\title{Emergence and Manipulation of non-equilibrium YSR states \\
\normalsize Supplemental Information}

\author{Jasmin Bedow, Eric Mascot and Dirk K. Morr}
\affiliation{University of Illinois at Chicago, Chicago, IL 60607, USA}

\maketitle

\section{Computation of Green's functions in the interaction representation of the Keldysh formalism}

The current flowing between an STM tip and a site in the superconductor is given by \cite{Keldysh1965,Caroli1971,Rammer1986}
\begin{equation}
  I(V,t) = \frac{e}{\hbar} (-t_\text{tun}) \sum_\sigma \left[ G^{<}_{ts} (\sigma,t,t) - G^{<}_{st} (\sigma, t,t) \right] \; ,
  \label{eq:I}
\end{equation}
where $t_{\text{tun}}$ is the tunneling amplitude between the tip and the superconductor, $V$ is the applied bias, and $G_{ts}^{<}$ is the equal-time lesser Green's function between the tip ($t$) and the site in the superconductor ($s$) that the electrons tunnel into \cite{Caroli1971}. To compute $G^{<}_{ts}$, we employ the Keldysh Dyson equations in real time, given by
\begin{align}
\bar{G}^{<}_{ts}(t,t) & = \bar{g}^{<}_{ts}(t,t) + \int dt_1 \bar{g}^{r}_{ts}(t,t_1) \bar{\mathcal{U}}(t_1) \bar{G}^{<}_{ss}(t_1,t) \nonumber \\
& + \int dt_1 \bar{g}^{<}_{ts}(t,t_1) \bar{\mathcal{U}}(t_1) \bar{G}^{a}_{ss}(t_1,t) \nonumber \\
\bar{G}^{a}_{ts}(t^\prime,t) & = \bar{g}^{a}_{ts}(t^\prime,t) + \int dt_1 \bar{g}^{a}_{ts}(t^\prime,t_1) \bar{\mathcal{U}}(t_1) \bar{G}^{a}_{ss}(t_1,t) \; ,
\label{eq:Dyson}
\end{align}
where $\bar{g}^{<,r,a}$ are the equilibrium lesser, retarded and advanced Greens function matrices of the unperturbed system in Nambu space, and $\bar{\mathcal{U}}(t)$ is the time-dependent perturbation matrix arising from Eq.(1) in the main text. To solve the above set of coupled equations, we rewrite the above integrals over time as sums of discrete time points $t_m = m \Delta t, m \in \{0,\ldots,N\}$, $t_0 = 0$ and $t_N = t$. The Dyson equations can then be rewritten as
\begin{align}
    \hat{G}^{<} (t_N,t_N) &= \hat{g}^{<} (t_N,t_N) + \sum_{m=0}^N \Delta t \ \hat{g}^{r}(t_N,t_m) \,  \hat{\mathcal{U}}(t_m) \, \hat{G}^{<} (t_m,t_N) + \sum_{m=0}^N \Delta t \ \hat{g}^{<}(t_N,t_m) \, \hat{\mathcal{U}}(t_m) \, \hat{G}^{a} (t_m,t_N) \nonumber \\
    \hat{G}^{r}(t_l,t_n) &= \hat{g}^{r} (t_l,t_n) + \sum_{m=0}^N \Delta t \, \hat{g}^{r}(t_l,t_m) \, \hat{\mathcal{U}}(t_m) \, \hat{G}^{r} (t_m,t_n)
    \label{eq:dyson_discrete}
\end{align}
where $ \hat{G}$ denotes a Greens function matrix in real and Nambu space. Defining Green's function matrices in both real space and time via (with $x=r,a,<$)
\begin{align}
    \check{G}^x =
    \left(\begin{array}{cccc}
        \hat{G}^x (t_0,t_0) & \hat{G}^x (t_0,t_1) & \cdots & \hat{G}^x (t_0,t_N) \\
        \hat{G}^x (t_1,t_0) & \hat{G}^x (t_1,t_1) & \cdots & \hat{G}^x (t_1,t_N) \\
        \vdots & \ddots & \ddots & \vdots \\
        \hat{G}^x (t_N,t_0) & \cdots & & \hat{G}^x (t_N,t_N)
    \end{array}\right)
    \; ,
    \label{eq:gf_matrix}
\end{align}
Eq.~(\ref{eq:dyson_discrete}) becomes
\begin{align}
    \check{G}^{<} &= \check{g}^{<} +  \Delta t \check{g}^{r} \check{\mathcal{U}} \check{G}^{<} + \Delta t \check{g}^{<} \check{\mathcal{U}} \check{G}^{a} \nonumber \\
    \check{G}^{r} &= \check{g}^{r} + \Delta t \check{g}^{r} \check{\mathcal{U}} \check{G}^{r} \label{eq:Gr}\; .
\end{align}
The solution of these equations is given by
\begin{align}
    \check{G}^{<} &= \left[1- \check{g}^{r} \check{\mathcal{U}} \, \Delta t \right]^{-1} \check{g}^{<} \left[1- \check{\mathcal{U}} \check{g}^a \, \Delta t \right]^{-1} \nonumber \\
    \check{G}^{r} &= \check{g}^{r} \left[1-  \check{\mathcal{U}} \check{g}^{r} \, \Delta t\right]^{-1} \; ,
\end{align}
where the diagonal elements of $\check{G}^{<}$ are the equal-time Greens functions that enter Eq.(\ref{eq:I}). The time dependence of the perturbation is fully encoded in the matrix $\check{\mathcal{U}}$, such that perturbations of arbitrary time-dependence can be considered.

To obtain the Greens function matrices $\hat{g}^{r} (t_l,t_n)$ of the unperturbed (equilibrium) system, we start by rewriting the unperturbed  Hamiltonian of the superconductor in Eq.(1) of the main text in Nambu space as
\begin{equation}
    \mathcal{H}_0 = \sum_{\bf k}
    \begin{pmatrix}
        c_{{\bf k},\uparrow}^\dagger & c_{-{\bf k},\downarrow}
    \end{pmatrix}
    \begin{pmatrix}
        \varepsilon_{\bf k} & \Delta_0 \\
        \Delta_0        & -\varepsilon_{-{\bf k}}
    \end{pmatrix}
    \begin{pmatrix}
        c_{{\bf k},\uparrow} \\
        c_{-{\bf k},\downarrow}^\dagger
    \end{pmatrix} = \sum_{\bf k} \psi_{\bf k}^\dagger {\bar H}_0 (\bf k) \psi_{\bf k}
\end{equation}
where $\varepsilon_k = - 2t_e (\cos{k_x} + \cos{k_y}) -\mu $. The corresponding retarded Greens function matrix is then obtained via
\begin{align}
    \bar{g}_{ss}^r ({\bf k},\omega) = \left[ \left( \omega + \mathrm{i}\Gamma \right) {\bar 1} - {\bar H}_0 \right]^{-1}
 \end{align}
where $\bar{1}$ is the $(2 \times 2)$ identity matrix, and  $\tau_\Gamma = \hbar/\Gamma$ is the quasi-particle lifetime. The retarded Greens function matrix in real space at a given site ${\bf r}$ is then given by
\begin{align}
    \bar{g}_{ss}^r (\omega) = \bar{g}_{ss}^r ({\bf r},{\bf r},\omega) = \int \frac{d^2k}{(2\pi)^2} \bar{g}_{ss}^r({\bf k},\omega) \; .
\end{align}
We next compute the Greens function of the superconductor, coupled to the STM tip via the tunneling amplitude $t_{tun}$, using
\begin{align}
    \left[ \hat{g}^r (\omega) \right]^{-1}= &
        \hat{g}_0^r (\omega)  - \hat{t}_\text{tun}
\end{align}
where
\begin{align}
\hat{g}_0^r (\omega) = \begin{pmatrix}
        \bar{g}_{tt}^r (\omega) & 0 \\
        0                         & \bar{g}_{ss}^r (\omega)
    \end{pmatrix} \ ; \qquad
    \hat{t}_\text{tun} = &
    -t_\text{tun}  \begin{pmatrix}
        0 & \bar{1} \\
        \bar{1} & 0
    \end{pmatrix}
    \; .
\end{align}
Here, $\bar{g}_{tt}^r$ is the retarded Greens function matrix of the tip, and $\hat{t}_\text{tun}$ is the tunneling matrix. For simplicity, we assume below that $\bar{g}_{tt}^r$ is equal to the normal state $\bar{g}_{ss}^r$, i.e., when $\Delta_0 = 0$. Similarly, the lesser Green's function matrix $\hat{g}^< (\omega)$ of the coupled system is given by
\begin{align}
    \hat{g}^< (\omega) &= \left[ \hat{1} - \hat{g}_0^r \hat{t}_\text{tun} \right]^{-1} \hat{g}_0^{<} \left[ \hat{1} - \hat{t}_\text{tun} \hat{g}_0^a \right]^{-1}
\end{align}
where
\begin{align}
    \hat{g}_0^{<}(\omega) &=
    \begin{pmatrix}
        -2\mathrm{i} \; \bar{n}_F^t(\omega) \; \mathrm{Im} \bar{g}^r_{tt} (\omega)  & 0 \\
        0                         & -2\mathrm{i} \; \bar{n}_F^s(\omega) \; \mathrm{Im} \bar{g}^r_{ss} (\omega)
    \end{pmatrix}
\end{align}
with the tip's and superconductor's Fermi distribution function matrices given by
\begin{equation}
    \bar{n}_F^t(\omega) = \begin{pmatrix}
        \frac{1}{\exp[\beta(\omega - e V)]+1} & 0 \\
        0 & \frac{1}{\exp[\beta(\omega + e V)]+1}
    \end{pmatrix}\ ; \qquad
    \bar{n}_F^s(\omega) = \begin{pmatrix}
        \frac{1}{\exp[\beta\omega]+1} & 0 \\
        0 & \frac{1}{\exp[\beta\omega]+1}
    \end{pmatrix} \; .
\end{equation}

In order to obtain the time-dependent equilibrium Greens functions that enter Eq.(\ref{eq:Dyson}),  we next Fourier transform the equilibrium Greens functions into time space via ($x=r,a,<$)
\begin{align}
    \hat{g}^{x}(t,t') = \hat{g}^{x}(t-t') = \int \frac{\mathrm{d}\omega}{2\pi} \hat{g}^{x}(\omega) e^{-\mathrm{i} \omega (t-t')} \; .
\end{align}

\section{Heisenberg representation of the Keldysh formalism for the calculation of Green's functions}

An alternative approach to describing the response of the system to a time-dependent perturbation is to compute the ``non-equilibrium'' density of states defined via
\begin{align}\label{eq:NEQLDOS}
N_{neq}({\bf r}, \sigma, t, \omega) & = -\frac{1}{\pi} {\rm Im} G^{r}({\bf r},{\bf r}, \sigma, t, \omega) \; .
\end{align}
Here, $G^{r}({\bf r},{\bf r}, \sigma, t, \omega)$ is obtained from the time-dependent retarded Greens function $G^{r}({\bf r},{\bf r},\sigma, t, t^\prime)$ by writing $t^\prime = t - \Delta t$ and performing a Fourier transform with regards to $\Delta t$, yielding
\begin{equation}
    G^{r}({\bf r},{\bf r}, \sigma, t, \omega) = \int_{-\infty}^{\infty} d(\Delta t) \, G^{r}({\bf r},{\bf r}, \sigma, t, t - \Delta t) \, e^{\mathrm{i} \omega \Delta t} \; .
    \label{eq:Gr_FT}
\end{equation}

To obtain $G^{r}({\bf r},{\bf r}, \sigma, t, t^\prime)$, we follow the approach of A. Kamanev and A. Levchenko \cite{Kamenev2009} and start from the generating functional given by
\begin{equation}
    Z = \frac{\mathrm{Tr}\left[ \rho_0 U(t_0,t) U(t,t_0)\right]}{\mathrm{Tr}\left[\rho_0\right]} \;,
    \label{eq:Z}
\end{equation}
which generates expectation values by functional differentiation.
Here, $\rho_0 = \rho(t_0)$ is the known density matrix of the system at time $t_0$ and $U(t,t_0)$ is the time-evolution operator from time $t_0$ to $t$, so that $\rho (t) = U(t,t_0) \rho_0 U(t_0,t)$ is the density matrix at time $t$. Expanding the numerator of Eq.(\ref{eq:Z}) yields
\begin{equation}
    \mathrm{Tr}\left[ \rho_0 U(t_0,t) U(t,t_0)\right] = \int D\left[\bar{\bm{\psi}},\bm{\psi}\right] \exp(i\bar{\bm{\psi}} \, \check{G}^{-1} \, \bm{\psi}) \; ,
\end{equation}
where the matrix $\check{G}$ contains blocks of time-ordered, lesser, greater and anti-time-ordered Green's function matrices and is given by
\begin{equation}
    \check{G} = \begin{pmatrix}
        \check{G}^T & \check{G}^{<} \\
        \check{G}^{>} & \check{G}^{\bar{T}}
    \end{pmatrix} \; .
\end{equation}
Each of these blocks is a matrix in both time, real and Nambu space, similar to Eq.~(\ref{eq:gf_matrix}).
The retarded Green's function is then obtained using
\begin{equation}
    \hat{G}^r (t,t') = \frac{1}{2} \left(\hat{G}^T (t,t') - \hat{G}^{\bar{T}} (t,t') + \hat{G}^{>} (t,t') - \hat{G}^{<} (t,t') \right) \; ,
\end{equation}
which yields
\begin{align}
    \hat{G}^r(t,t') &= - \mathrm{i} \Theta(t-t') \hat{S}(t,t') \ ,
\end{align}
with
\begin{align}
    \hat{S}(t,t') & = T \exp\left( -\frac{\mathrm{i}}{\hbar} \int_{t'}^{t} \hat{H}(t_1) d{t_1} \right) ,
    \label{eq:S_matrix}
\end{align}
where $\hat{G}(t,t')$ and $\hat{S}(t,t')$ are matrices in real and Nambu space.
Here, $\hat{H}$ is the matrix representation of the Hamiltonian in Eq.(1) of the main text obtained via
\begin{align}
    H(t) = \psi^\dagger(t) \hat{H} (t) \psi(t) \;
\end{align}
where $\psi^\dagger,\psi$ are spinors in real and Nambu space,
and $T$ is the time-ordering operator. This operator ensures that in the expansion of the exponential in Eq.~(\ref{eq:S_matrix}), the matrices are time-ordered. Moreover, we assume for concreteness that the perturbation starts at a time $t=0$, allowing us to write the  time-dependent Hamiltonian as
\begin{equation}
    \hat{H}(t) = \Theta(-t) \hat{H}_0 + \Theta(t) \left[ \hat{H}_0 + \mathcal{\hat{U}}(t) \right]
\end{equation}
where $\Theta(t)$ is the Heaviside $\Theta$-function.
Using
\begin{equation}
    \frac{d}{dt} \hat{S}(t,t') = - \mathrm{i} \hat{H}(t) \hat{S}(t,t') \; ,
\end{equation}
which immediately follows from Eq.~(\ref{eq:S_matrix}), and taking the derivative with respect to $t$ of Eq.~(\ref{eq:Gr_FT}), we then obtain the differential equation
\begin{equation}
    \left[ \mathrm{i} \frac{d}{dt} +  \omega  + i \Gamma - {\hat H}(t)\right] {\hat G}^r \left(t, \omega \right) = {\hat 1} \; .
\end{equation}
This equation was previously derived in the context of the Floquet formalism \cite{Kohler2005,Liu2019}.
Note that for $t \leq 0$, $\frac{d}{dt} \hat{G}^r(t,\omega) = 0$ since $\hat{H}(t<0) = \hat{H}_0$, and we thus obtain for $\hat{G}^r(t<0,\omega)$ simply the retarded Green's function of the unperturbed system $\hat{g}^r_0(\omega)$. Thus, $\hat{G}^r(t=0,\omega) = \hat{g}_0^r(\omega)$ can be used as an initial condition to solve the differential equation numerically by integrating forward in time.

For the special case that the Hamiltonian's time-dependence is determined solely through a Heaviside-function, i.e.
\begin{equation}
    \hat{H}(t) = \Theta(-t) \hat{H}_0 + \Theta(t) \, \hat{H}^{+} \; ,
\end{equation}
we can obtain an explicit form of the Green's function by carrying out the integration in Eq.~(\ref{eq:Gr_FT}) analytically, which yields
\begin{equation}
    \hat{G}^r(t,\omega) = \left[]1 - e^{\mathrm{i} \left( \omega \hat{1} - \hat{H}^{+} \right) t} e^{ - \Gamma t} \right] \hat{g}^r_+(\omega) + e^{\mathrm{i} \left( \omega \hat{1} - \hat{H}^{+} \right) t} e^{ - \Gamma t} \hat{g}^r_0(\omega) \; ,
\end{equation}
where the equilibrium Greens function are given by
\begin{align}
    \hat{g}^r_{0}(\omega) = \frac{1}{\omega - \hat{H}_{0} + \mathrm{i} \Gamma} \; ; \qquad
    \hat{g}^r_{+}(\omega) = \frac{1}{\omega - \hat{H}^{+} + \mathrm{i} \Gamma} \; .
\end{align}
Note that $\hat{G}^r(t \rightarrow \infty,\omega) = \hat{g}^r_{+}(\omega)$.

\section{Floquet formalism for perturbations periodic in time}

If the system is described by a Hamiltonian which is periodic in time $\mathcal{H}(t) = \mathcal{H}(t+T)$ with a period $T$, we can employ the Floquet formalism \cite{Kohler2005,Sacramento2015,Liu2019} to compute the time-dependent Greens function. Writing the Hamiltonian as a matrix in real space, we can expand its time-dependent and periodic (in time) eigenfunctions in a Fourier series. This yields the infinite-dimensional Floquet-Sambe Hamiltonian matrix
\begin{equation}
   \hat{H}_{FS} =  \left(
    \begin{array}{ccccccc}
        \cdots & \cdots & \cdots & \cdots & \cdots & \cdots & \cdots \\
        \cdots & {\hat H}_{-2,-2} & {\hat H}_{-2,-1} & {\hat H}_{-2,0} & {\hat H}_{-2,1} & {\hat H}_{-2,2} & \cdots \\
        \cdots & {\hat H}_{-1,-2} & {\hat H}_{-1,-1} & {\hat H}_{-1,0} & {\hat H}_{-1,1} & {\hat H}_{-1,2} & \cdots \\
        \cdots & {\hat H}_{0,-2} & {\hat H}_{0,-1} & {\hat H}_{0,0} & {\hat H}_{0,1} & {\hat H}_{0,2} & \cdots \\
        \cdots & {\hat H}_{1,-2} & {\hat H}_{1,-1} & {\hat H}_{1,0} & {\hat H}_{1,1} & {\hat H}_{1,2} & \cdots \\
        \cdots & {\hat H}_{2,-2} & {\hat H}_{2,-1} & {\hat H}_{2,0} & {\hat H}_{2,1} & {\hat H}_{2,2} & \cdots \\
        \cdots & \cdots & \cdots & \cdots & \cdots & \cdots & \cdots \\
    \end{array}
    \right)
    \; .
\end{equation}
with block elements
\begin{equation}
    {\hat H}_{m,m'} = m \omega_0 {\hat 1} \delta_{m,m'} + \frac{1}{T} \int_0^T dt e^{-\mathrm{i} m \omega_0 t} {\hat H}(t) e^{\mathrm{i} m' \omega_0 t } \;,
\end{equation}
where $\omega_0 = 2\pi/T$. Here, ${\hat H}_{0,0}$ describes the unperturbed, static part of the Hamiltonian. To analyze this matrix, one renders it finite-dimensional by choosing a cut-off index $l$ and retaining only those block elements $\hat{H}_{m,m^\prime}$ with $ -l \leq m,m' \leq l$. \\

In the main text, we consider the case of a single impurity located at ${\bf r}=0$ whose spin periodically rotates in the $xz$-plane. The Hamiltonian matrix in Nambu space then takes the form
\begin{equation}
    \hat{H}(t) = \hat{H}_0 + JS \hat{\delta}_{{\bf r},0} \cdot \left[\cos(\omega_0 t) \, \tau_0 \otimes \sigma_z + \sin(\omega_0 t) \, \tau_0 \otimes \sigma_x\right]
\end{equation}
where ${\hat H}_0$ is the static part of the Hamiltonian matrix in real space, and  $\hat{\delta}_{{\bf r},0}$ is a real-space matrix with unity elements at the site of the impurity ${\bf r}=0$ only. This yields for the Floquet-Sambe Hamiltonian matrix (shown around $m,m'=0$)
\begin{equation}
   H_{FS}  =  \left(
    \begin{array}{ccccccc}
        \cdots & \cdots & \cdots & \cdots &  \cdots & \cdots & \cdots \\
        \cdots & \hat{J}^* & -\omega_0 \hat{1} + \hat{H}_0  & \hat{J} & 0 & 0 & \cdots \\
        \cdots & 0 & \hat{J}^* & \hat{H}_0  & \hat{J} & 0 &  \cdots \\
        \cdots & 0 & 0 & \hat{J}^* & \omega_0 \hat{1} + \hat{H}_0  & \hat{J}& \cdots \\
        \cdots & \cdots & \cdots & \cdots & \cdots & \cdots & \cdots \\
    \end{array}
    \right)
    \; .
\end{equation}
where
\begin{align}
    \hat{J} =\frac{1}{2} \hat{\delta}_{{\bf r},0} \tau_0 \otimes (\sigma_z + \mathrm{i} \sigma_x) \ .
\end{align}

Denoting the eigenvalues of $H_{FS}$ by $E_n$, we can write the corresponding eigenvectors in the form
\begin{equation}
    \hat{\Phi}_n = \left(
    \begin{array}{c}
    \vdots \\
    {\hat \phi}_n^{(m=-1)} \\
       {\hat \phi}_n^{(m=0)} \\
       {\hat \phi}_n^{(m=1)} \\
       \vdots
    \end{array}
    \right)
\end{equation}
where $\phi_n^{(m)}$ is a column vector of length $4N$, where $N$ is the number of sites in the system and the factor of 4 appears due to the dimension of Nambu space. The retarded Green's function matrix $\hat{G}^r(\omega,t)$ in real space can then be written as a Fourier expansion in the frequency $\omega_0$ \cite{Kohler2005,Liu2019} and is given by
\begin{equation}
    {\hat G}^r(\omega,t) = \sum_{p = -\infty}^\infty {\hat G}^{r}_{p} (\omega) e^{\mathrm{i} p \omega_0 t}
\end{equation}
where
\begin{equation}
    {\hat G}^{r}_p (\omega) = \sum_{n} \sum_{m = -\infty}^\infty \frac{{\hat \phi}_n^{p+m} \left[{\hat \phi}^{m}_n\right]^T}{\omega - E_n + m \omega_0 + i \Gamma} \; .
\end{equation}

\section{Comparison of $N_{Fl}$ and $N_{neq}$ for a periodically rotating spin}

In Fig.~4 of the main text, we discussed the form of $N_{Fl}$ and $N_{neq}$ for a periodically rotating spin in the $xz$-plane. A comparison of $N_{Fl}$ and $N_{neq}$ at $t=400 \tau_e$ shown in Supplementary Fig.~\ref{fig:SI_Fig1} shows good quantitative agreement implying that $N_{neq}$ is suited to describe not only the transition from a static to a periodically driven system, but also a periodically driven system itself.
\begin{figure}
  \centering
  \includegraphics[width=10cm]{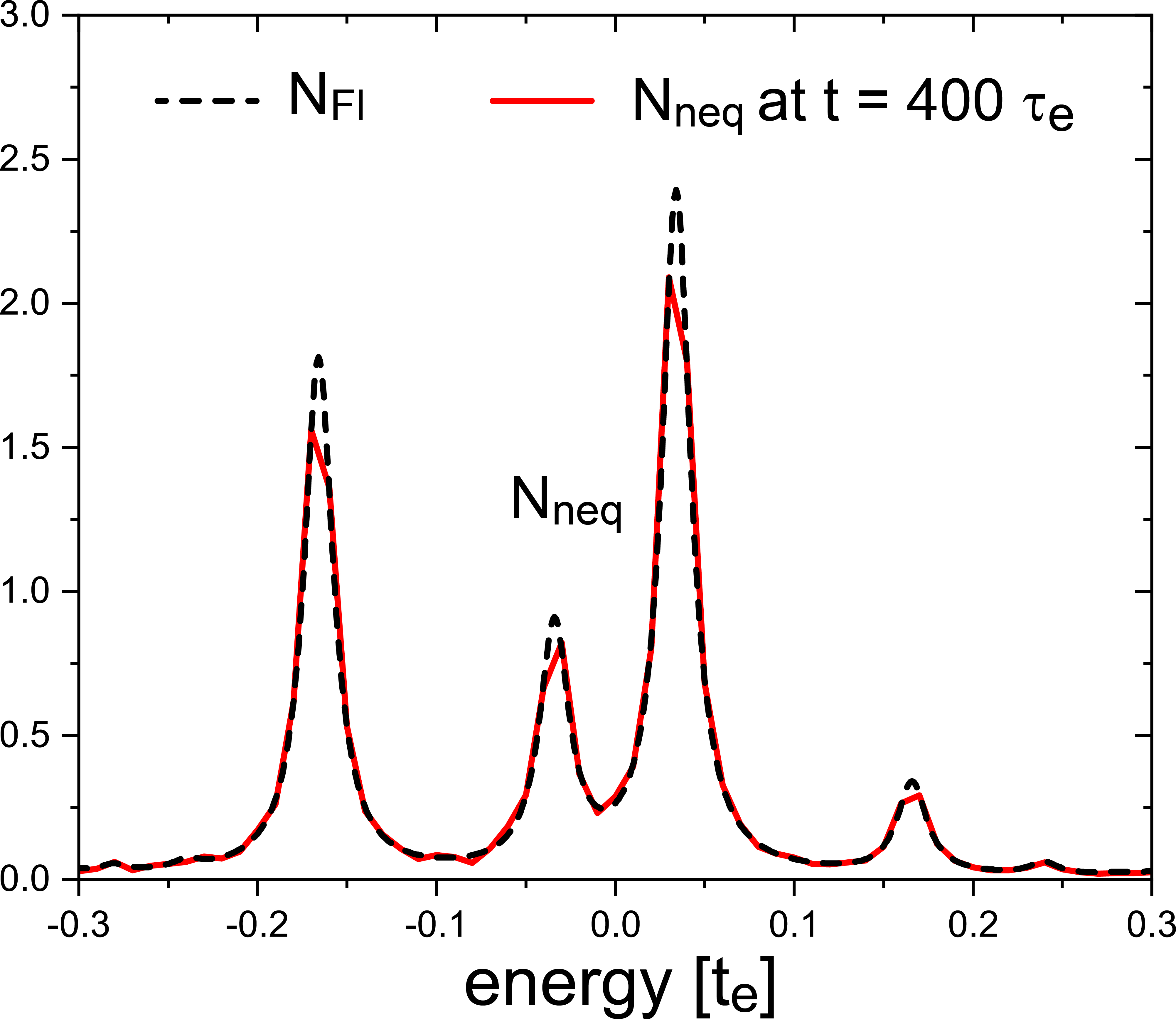}
  \caption{ Comparison of $N_{Fl}$ and $N_{neq}$ at $t=400 \tau_e$ for a periodically rotating spin, as discussed in Fig.4 of the main text.}
  \label{fig:SI_Fig1}
\end{figure}\\